\documentclass[reprint,aps,onecolumn,11pt,notitlepage]{revtex4-1}
\usepackage[english]{babel}
\usepackage{amsmath,amssymb,graphicx,bm}
\usepackage[colorlinks=true, linkcolor=red, citecolor=blue]{hyperref}
\usepackage[justification=raggedright,font=small,labelfont=bf]{caption}

\begin{document}
\title{Rotating and twisting charged black holes with cloud of strings and quintessence as a particle accelerator}

\author{Qi-Quan Li}
\author{Yu Zhang}
\email{zhangyu\_128@126.com}
\author{Qi Sun}
\author{Chen-Hao Xie}
\affiliation{Faculty of Science, Kunming University of Science and Technology, Kunming, Yunnan 650500, China.}
\begin{abstract}
In this paper, we study the effects of the rotation parameter $a$, the twist parameter $n$, the string cloud parameter $b$, the quintessence state parameter $\omega _{q}$ and the charge parameter $q$ on the horizons and ergosphere of rotating and twisting charged black holes with cloud of strings and quintessence, and obtain the equations of motion and effective potential of the particle on the equatorial plane of black hole. We find that a particle with the critical angular momentum $L = L_C$ falling from infinity reaches the event horizon($u^r$=0) and satisfies the circular orbit condition $V_{e f f}={V_{e f f}}'=0$. We derive the expression of the centre-of-mass (CM) energy of two particles with different masses from the equations of particle motion. We show that the CM energy can be arbitrarily large for extremal black holes when the particles reach the event horizon by adjusting the angular momentum of the incident particles. However, for non-extremal black holes, the CM energy of particles that reach the event horizon cannot diverge by adjusting the angular momentum of the incident particles.
\end{abstract}

\maketitle 

\section{Introduction}
It is a well-known fact that our universe is expanding at an accelerated rate \cite{SupernovaCosmologyProject:1998vns, SupernovaSearchTeam:2004lze}, and one of the possible explanations for this phenomenon is the existence of dark energy. Dark energy is an unknown form of energy that is presumed to exist in the universe and accounts for more than 70$\%$ of the total energy in the observable universe. The quintessence \cite{Kiselev:2002dx} is a hypothetical form of dark energy based on a dynamic scalar field, whose equation of state $p_{q}=w_{q} \rho_{q}$ describes the relationship between the pressure $p_{q}$ and the energy density $\rho_{q}$. 
When $-1<w_{q}<-1/3$, the quintessence creates a negative pressure that leads to the accelerated expansion of the universe.

String theory explains how fundamental unit of nature is one-dimensional strings, which vibrate in different modes, producing various elementary particles and interaction forces. The extension of this idea is to consider a cloud of strings and study the possible measurable effects it has on the black hole gravitational field. The earliest theoretical analysis of these cloud of strings was done by Letelier \cite{Letelier:1979ej} and obtained the generalization of the Schwarzschild solution corresponding to the black hole surrounded by a cloud of strings, as well as some other interesting results \cite{Letelier:1981,Letelier:1983}.

The Taub-NUT spacetime \cite{Misner:1963fr} was proposed by Misner after considering Taub's work in 1951 \cite{Taub:1950ez} and Newman, Unti and Tamburino's work in 1963 \cite{Newman:1963yy}. Its metric form is as follows:
\begin{equation}
	\begin{aligned}
		\begin{array}{c}
			d s^{2}=-f(r) d t^{2}+\frac{d r^{2}}{f(r)}+\left(r^{2}+n^{2}\right)\left(d \theta^{2}+\sin ^{2} \theta d \phi^{2}\right)+4 n r \sin ^{2} \frac{\theta}{2} d t d \phi,
		\end{array}
	\end{aligned}
\end{equation}
where
\begin{equation}
	\begin{aligned}
		\begin{array}{c}
			f(r)=\frac{r^{2}-2 M r-n^{2}}{r^{2}+n^{2}},
		\end{array}
	\end{aligned}
\end{equation}
$M$ and $n$ are positive constants. The understanding of the black hole parameter $n$ is that it not only represents the NUT charge, but also the twisting parameter. Then Carter suggested the Kerr-Taub-NUT black hole \cite{Carter:1966zza}, and Newman et al. proposed the Kerr-Newman-Taub-NUT black holes \cite{Newman:1963yy,Demia}. More kinds of black holes with NUT charge were also suggested \cite{Sorkin:1983ns,Brill:1997mf,Vacaru:2001ak,Dehghani:2006aa,Aliev:2008wv,Hendi:2008wq,Nedkova:2011aa,deMToledo:2018tjq,Sakti:2019iku}.

In 2009, Banados, Silk and West (BSW) \cite{Banados:2009pr} proposed that by adjusting the angular momentum of two particles on the equatorial plane in the background of an extreme Kerr black hole, the CM energy of the particles can diverge when they reach the event horizon. Subsequently, Mao et al. not only proved that a charged nonrotating Kaluza-Klein black hole \cite{Mao:2010di} can serve as a particle accelerator, but also proved that a large amount of energy can be obtained near a Kaluza-Klein naked singularity \cite{Mao:2010di}. It had also been demonstrated that various kinds of black holes could serve as particle accelerators, such as regular black holes (rotating Bardeen black hole \cite{Ghosh:2015pra}, rotating Hayward black hole \cite{Amir:2015pja}), black holes in higher dimensions (five-dimensional Myers-Perry black holes \cite{Abdujabbarov:2013qka}) and lower dimensions (charged dilaton black hole in 2+1 dimensions \cite{Sadeghi:2013gmf}, rotating charged hairy black hole in (2+1) dimensions \cite{Fernando:2017qrq}), and many other black holes \cite{Zaslavskii:2010jd,Tursunov:2013zha,Wei:2010vca,Abdujabbarov:2011af,Zhang:2018ocv,Patil:2011aa,Lake:2010bq,Liu:2010ja,Patil:2010nt,Patil:2011aw}.
The results of the above articles show that most rotating and charged black holes can be used as particle accelerators to make the CM energy of particles reach arbitrarily high levels. This has important implications for human to explore physics at the Planck energy scale and explanation of astrophysical phenomena.

In this paper, we mainly study rotating and twisting charged black holes with cloud of strings and quintessence as particle accelerators. The paper is organized as follows. In Sec.~\ref{II}, we review this black hole and study the effects of its various parameters on the horizon and the ergosphere. In Sec.~\ref{III}, we derive the motion equation and the effective potential of the particle on the black hole equatorial plane, and obtain the range of angular momentum that allows the particle to enter the black hole, and introduce the critical angular momentum. In Sec.~\ref{IV}, we derive the CM energy expression of two particles with different masses outside the black hole, and calculate the CM energy at the event horizon for extreme and non-extreme black holes. In Sec.~\ref{V}, we summarize the article.

\section{ROTATING AND TWISTING CHARGED BLACK HOLES WITH CLOUD OF STRINGS AND QUINTESSENCE} \label{II}
Recently, Sakti et al. obtained the solution of rotating and twisting charged black holes with cloud of strings and quintessence \cite{Sakti:2019iku}. Let's now review the black hole solution. For this black hole solution, we need to first solve the non-zero components of the energy-momentum tensor of three kinds of matter: the electromagnetic field, a cloud of strings and the quintessence.

The non-zero components of the energy-momentum tensor of the electromagnetic field are \cite{Sakti:2019krw}
\begin{equation}\label{eq:EM1}
	\begin{aligned}
		\begin{array}{c}
			E_{t}^{t}=E_{r}^{r}=-\frac{e^{2}+g^{2}}{r^{4}}, \quad E_{\theta}^{\theta}=E_{\phi}^{\phi}=\frac{e^{2}+g^{2}}{r^{4}},
		\end{array}
	\end{aligned}
\end{equation}
where $e$ and $g$ are the electric and magnetic charges, respectively.
The non-zero components of the energy-momentum tensor of the cloud of strings are \cite{Letelier:1979ej}
\begin{equation}\label{eq:EM2}
	\begin{aligned}
		\begin{array}{c}
			B_{t}^{t}=B_{r}^{r}=-\frac{b}{r^{2}}, \quad B_{\theta}^{\theta}=B_{\phi}^{\phi}=0,
		\end{array}
	\end{aligned}
\end{equation}
where $b$ is related to the strength of a cloud of strings.
The non-zero components of the energy-momentum tensor of the quintessence are \cite{Kiselev:2002dx}
\begin{equation}\label{eq:EM3}
	\begin{aligned}
		\begin{array}{c}
			Q_{t}^{t}=Q_{r}^{r}=-\rho_{q}(r), \quad Q_{\theta}^{\theta}=Q_{\phi}^{\phi}=\frac{1}{2} \rho_{q}(r)\left(3 \omega_{q}+1\right),
		\end{array}
	\end{aligned}
\end{equation}
where $\omega _{q}$ and $\rho _{q}$ are the equation of state parameter and energy density, respectively.
In the static spherically symmetric spacetime background, the metric has the following form:
\begin{equation}\label{eq:ds2}
	\begin{aligned}
		\begin{array}{c}
			\mathrm{d} s^{2}=-f(r) \mathrm{d} t^{2}+f(r)^{-1} \mathrm{~d} r^{2}+r^{2} \mathrm{~d} \Omega^{2},\\
			\mathrm{~d} \Omega^{2}=d \theta^{2}+\sin ^{2} \theta d \phi^{2}.
		\end{array}
	\end{aligned}
\end{equation}
The result obtained by substituting Eqs.(\ref{eq:EM1}), (\ref{eq:EM2}), (\ref{eq:EM3}), and (\ref{eq:ds2}) into the Einstein field equation $G_{\mu\nu}=T_{\mu\nu}$ is
\begin{equation}\label{eq:ds3}
	\begin{aligned}
		\begin{array}{c}
			f(r)=1-\frac{2 M}{r}+\frac{q^{2}}{r^{2}}-\alpha r^{-1-3 \omega_{q}}-b,
		\end{array}
	\end{aligned}
\end{equation}
where $q^2=e^2+g^2$, $M$ and $\alpha$ are the black hole mass and quintessential intensity, respectively.
The DJN algorithm for solving the rotating and twisting black hole solutions was first proposed by Demia\'nski, which use the tetrad formalism \cite{Demianski:1972uza}. In 2016, Erbin simplified the DJN algorithm by using some simple coordinate transformations \cite{Erbin:2014aya,Erbin:2014aja}. In Ref. \cite{Sakti:2019iku}, the metric for the rotating and twisting charged black holes with cloud of strings and quintessence was obtain by the DJN algorithm in the following form:
\begin{equation}\label{eq:ds4}
	\begin{aligned}
		\begin{array}{c}
			\begin{aligned}d s^{2}= & -\frac{\Delta}{\rho^{2}}\left[d t-\left\{a \sin ^{2} \theta+2 n(1-\cos \theta)\right\} d \varphi\right]^{2}+\frac{\rho^{2}}{\Delta} d r^{2} \\& +\rho^{2} d \theta^{2}+\frac{\sin ^{2} \theta}{\rho^{2}}\left[a d t-\left\{r^{2}+(a+n)^{2}\right\} d \varphi\right]^{2},\end{aligned}
		\end{array}
	\end{aligned}
\end{equation}
with
\begin{equation}\label{eq:ds5}
	\begin{aligned}
		\begin{array}{c}
			\rho^{2}=r^{2}+(n+a \cos \theta)^{2}, \\ \Delta=r^{2}-2 M r+a^{2}+q^{2}-n^{2}-\alpha r^{v}-b r^{2},\\ v=1-3 \omega_{q}.
		\end{array}
	\end{aligned}
\end{equation}
The two metrics are similar when $b$=0 in Eq.(\ref{eq:ds4}) and $\kappa \lambda$=0 in the Kerr-Newman-NUT-Kiselev black hole solution in Rastall theory of gravity \cite{Sakti:2019krw}.

\subsection{Horizons and ergosphere}

\begin{figure*}
	\begin{tabular}{c c}
		\includegraphics[scale=0.7]{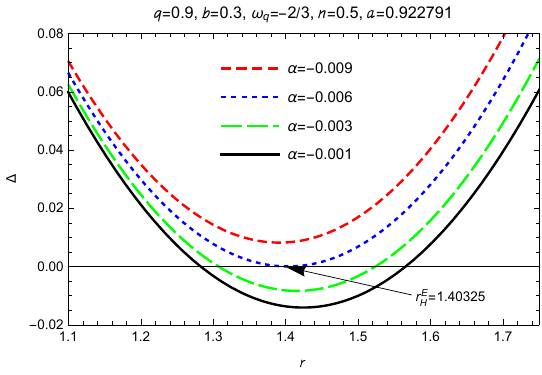}\hspace{-0.2cm}
		&\includegraphics[scale=0.7]{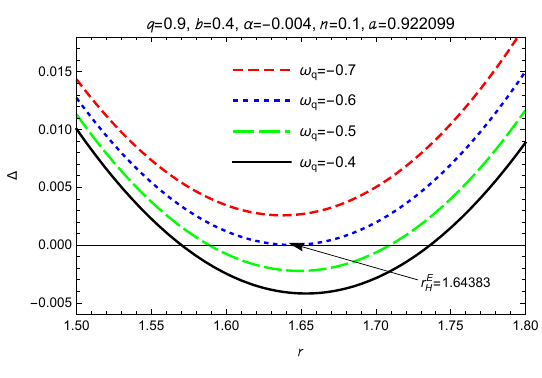}\\
		\includegraphics[scale=0.7]{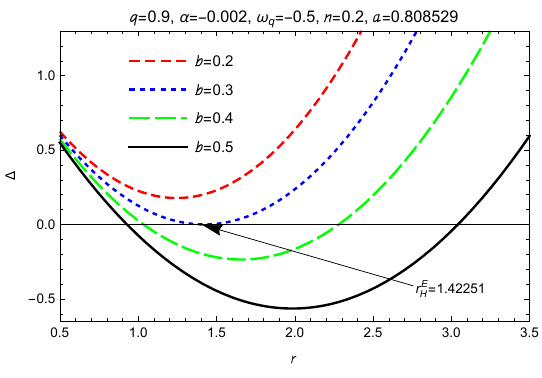}\hspace{-0.2cm}
		&\includegraphics[scale=0.7]{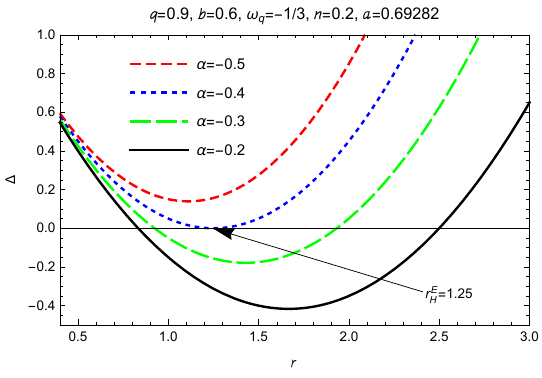}
	\end{tabular}
	\caption{Graphs of $\Delta$ vs $r$ for different $\alpha$, $\omega _{q}$ and $b$ values under fixed $q$, $n$ and $a$. The blue dashed line corresponds to the extreme black hole.}\label{fig:horizon2}
\end{figure*}

In the black hole metric (\ref{eq:ds4}), the expression for the black hole horizon is obtained by setting $g^{rr}$=0, which is to solve
\begin{equation}\label{eq:HO}
	\begin{aligned}
		\begin{array}{c}
			\Delta=r^2-2 M r+a^2+q^2-n^2-\alpha r^{1-3 w}-b r^2=0 .
		\end{array}
	\end{aligned}
\end{equation}
\begin{figure*}
	\begin{tabular}{c c}
		\includegraphics[scale=0.7]{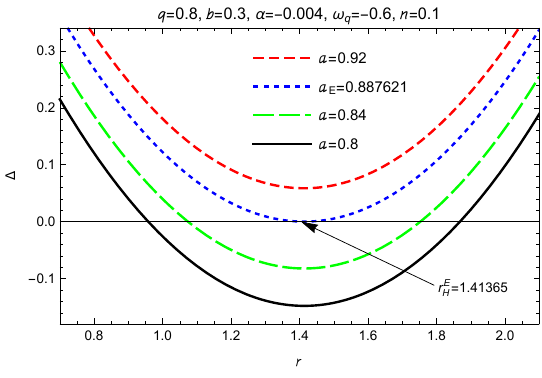}\hspace{-0.2cm}
		&\includegraphics[scale=0.7]{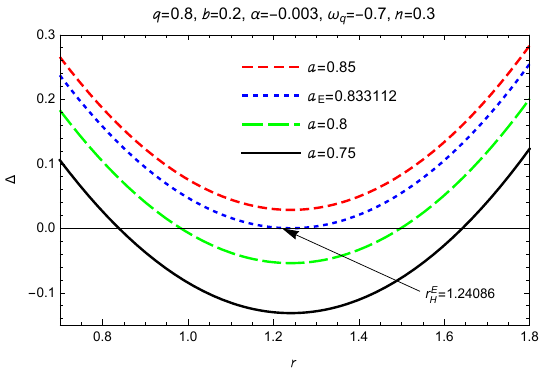}\\
		\includegraphics[scale=0.7]{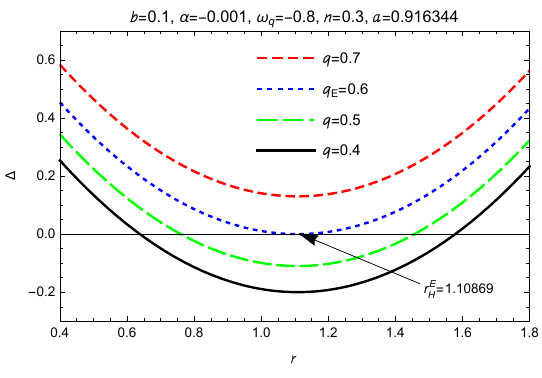}\hspace{-0.2cm}
		&\includegraphics[scale=0.7]{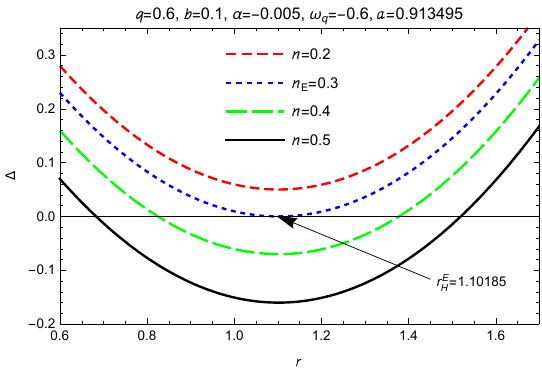}
	\end{tabular}
	\caption{Graphs of $\Delta$ vs $r$ for different $q$, $n$ and $a$ values under fixed $b$, $\alpha$ and $\omega _{q}$. The blue dashed line corresponds to the extreme black hole.}\label{fig:horizon1}
\end{figure*}
By numerical analysis of Eq.(\ref{eq:HO}), we find that rotating and twisting charged black holes with cloud of strings and quintessence may have two roots, which are called the Cauchy horizon $r_{H}^{-}$ and the event horizon $r_{H}^{+}$, respectively. The behavior of the event horizon for different parameters $q$, $b$, $\alpha$, $\omega _{q}$, $n$ and $a$ is shown in Figs.~\ref{fig:horizon1} and \ref{fig:horizon2}. From Figs.~\ref{fig:horizon1} and \ref{fig:horizon2}, we can draw the following conclusions:

\begin{itemize}
	\item{From Fig.~\ref{fig:horizon1}, we can find that by changing the parameters $q$, $a$ or $n$ respectively when the other parameters are determined, we can get critical value $q_E$, $a_E$ or $n_E$ which makes the inner and outer horizons of the black hole coincide.}	
	\item{From the first three figures in Fig.~\ref{fig:horizon1}, we find that when other parameters remain unchanged, the distance between the inner and outer horizons of the black hole decreases with the increase of the values of the parameters $q$ and $a$, and when the parameter $q$ or $a$ is a critical value, the inner and outer horizons coincide; when the value of $q$ or $a$ is greater than the critical value, there is no black hole.}
	\item{From the last figure in Fig.~\ref{fig:horizon1}, we can find that when other parameters remain unchanged, the distance between the inner and outer event horizons of the black hole decreases with the decrease of the value of parameter $n$, and when the parameter $n$ is a critical value, the inner and outer horizons coincide; when the value $n$ is less than the critical value, there is no black hole.}
	\item{From the first three figures in Fig.~\ref{fig:horizon2}, we find that the minimum value of the $\Delta$ function decreases as the parameters $\alpha$, $\omega _{q}$ and $b$ increase, and the radial coordinate $r_{min}$ corresponding to the minimum value of the $\Delta$ function also decreases as the parameters increase. From the third and fourth figures in Fig.~\ref{fig:horizon2}, we find that the parameter $\alpha$ in the fourth figure when $\omega _{q}=-1/3$ and the parameter $b$ in the third figure when $\omega _{q}$ ($-1<\omega _{q}<-1/3$) is equal to other values have similar effects on the $\Delta$ function. The reason is that both parameter $\alpha$ when $\omega _{q}=-1/3$ and parameter $b$ when $\omega _{q}$ is equal to other values affect the $\Delta$ function by changing the coefficients in front of the $r^2$ term.
	}	
	
\end{itemize}

\begin{figure*}
	\begin{center}
		\begin{tabular}{c c c c}
			\includegraphics[scale=0.48]{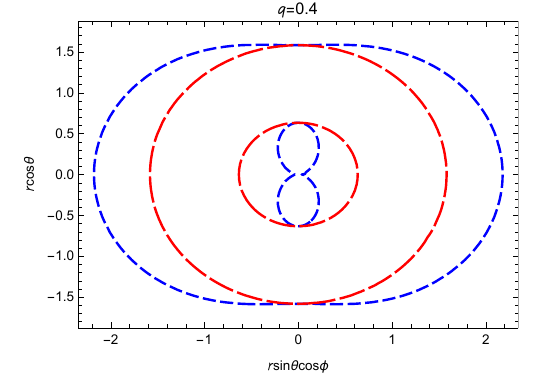}\hspace{-0.45cm}
			\includegraphics[scale=0.48]{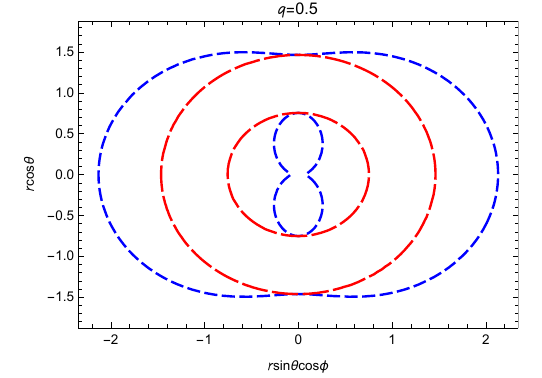}\hspace{-0.45cm}
			\includegraphics[scale=0.48]{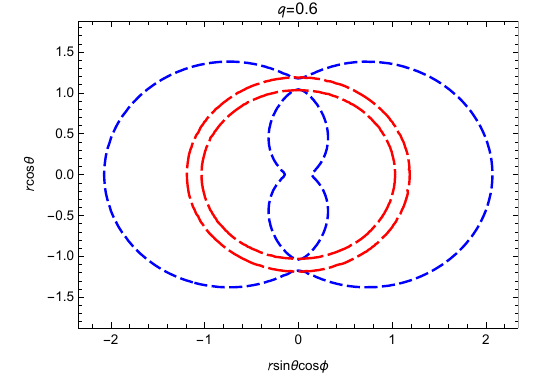}\hspace{-0.7cm}
			&\includegraphics[scale=0.48]{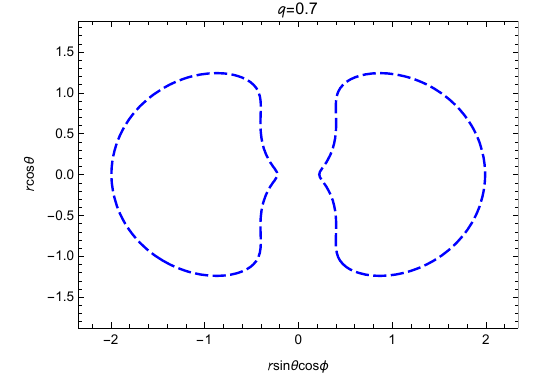}\\
			\includegraphics[scale=0.48]{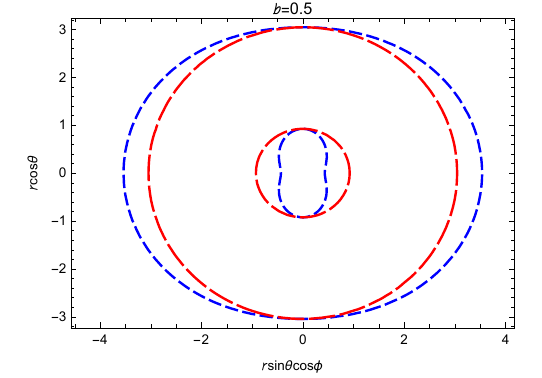}\hspace{-0.45cm}
			\includegraphics[scale=0.48]{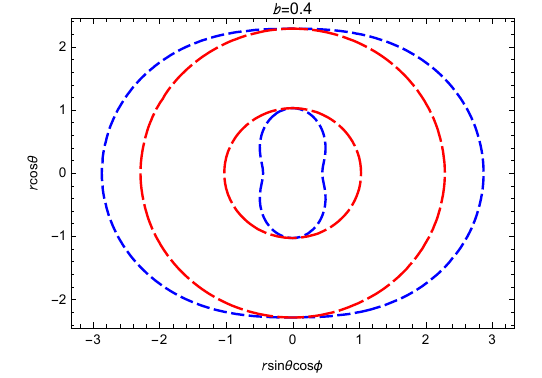}\hspace{-0.45cm}
			\includegraphics[scale=0.48]{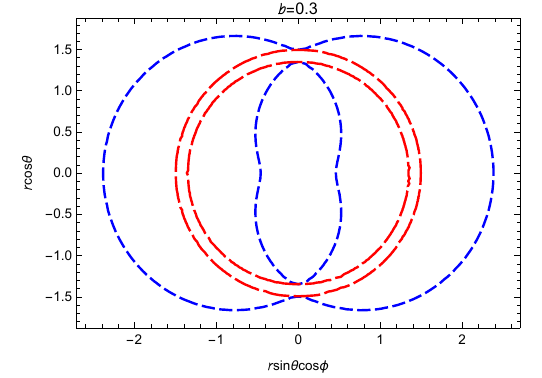}\hspace{-0.7cm}
			&\includegraphics[scale=0.48]{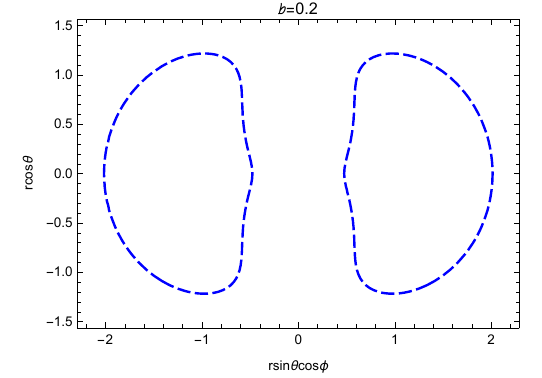}\\
			\includegraphics[scale=0.48]{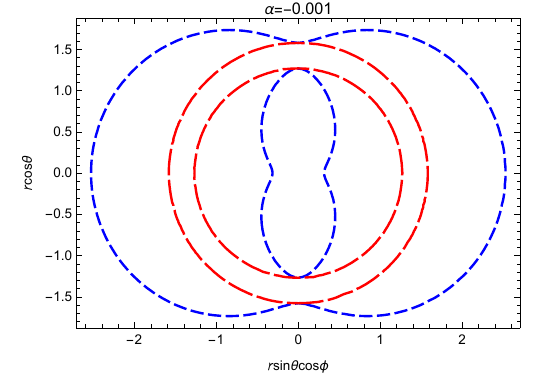}\hspace{-0.45cm}
			\includegraphics[scale=0.48]{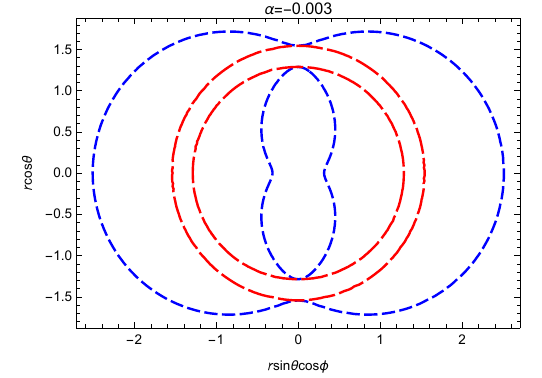}\hspace{-0.45cm}
			\includegraphics[scale=0.48]{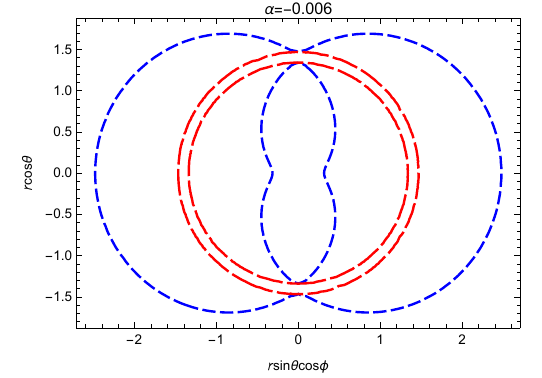}\hspace{-0.7cm}
			&\includegraphics[scale=0.48]{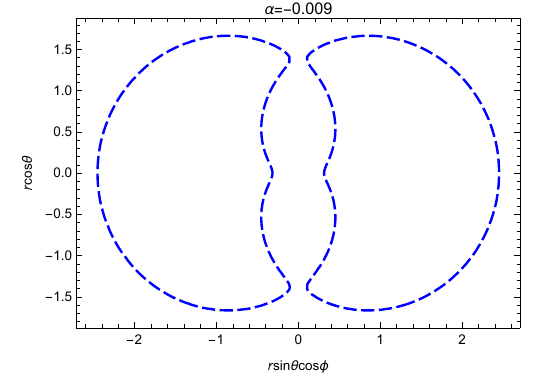}\\
			\includegraphics[scale=0.48]{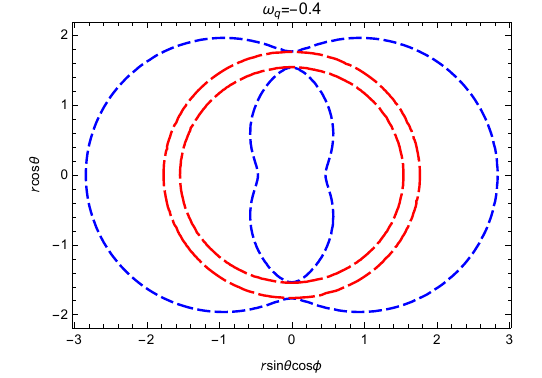}\hspace{-0.45cm}
			\includegraphics[scale=0.48]{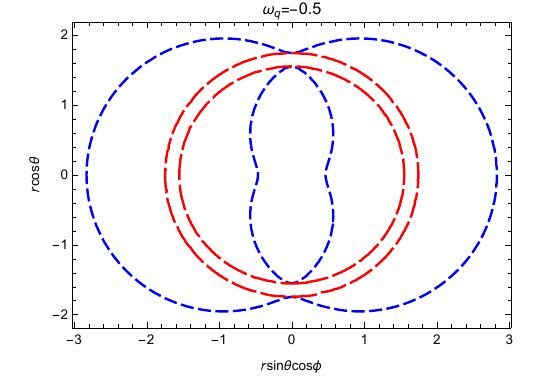}\hspace{-0.45cm}
			\includegraphics[scale=0.48]{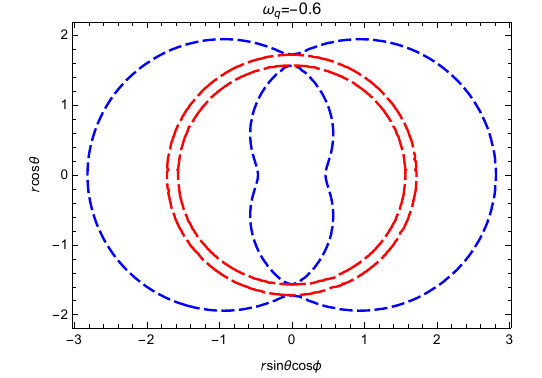}\hspace{-0.7cm}
			&\includegraphics[scale=0.48]{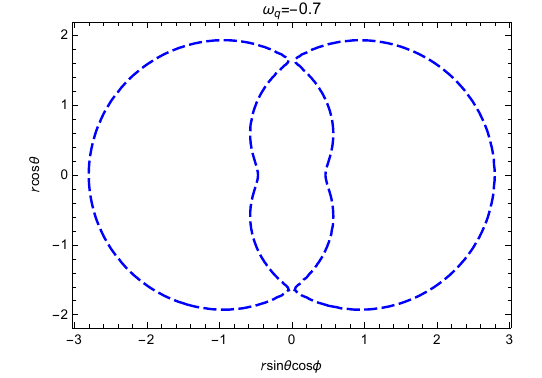}\\
			\includegraphics[scale=0.48]{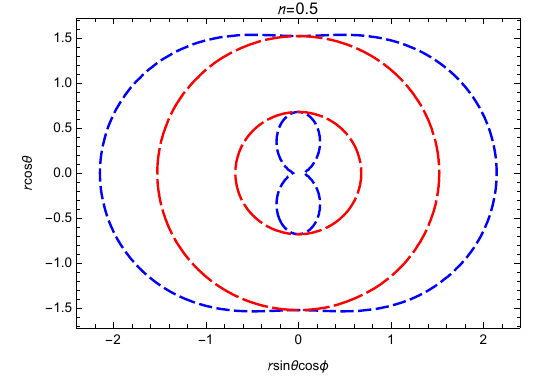}\hspace{-0.45cm}
			\includegraphics[scale=0.48]{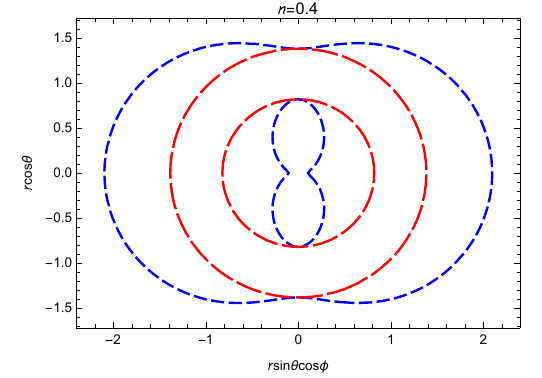}\hspace{-0.45cm}
			\includegraphics[scale=0.48]{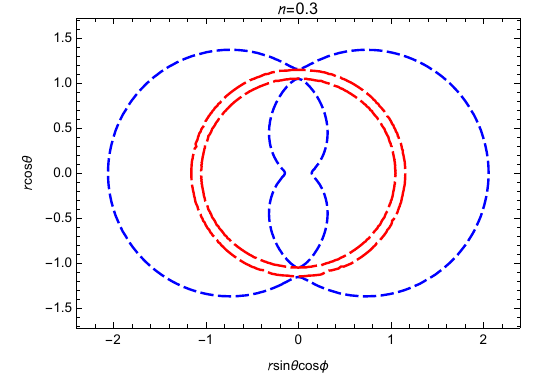}\hspace{-0.7cm}
			&\includegraphics[scale=0.48]{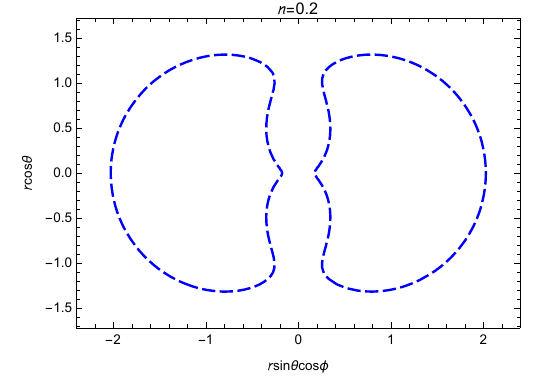}\\
			\includegraphics[scale=0.48]{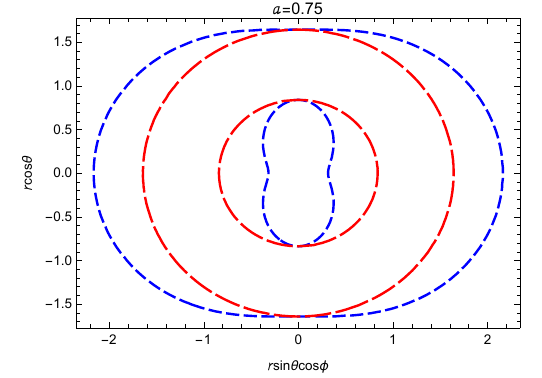}\hspace{-0.45cm}
			\includegraphics[scale=0.48]{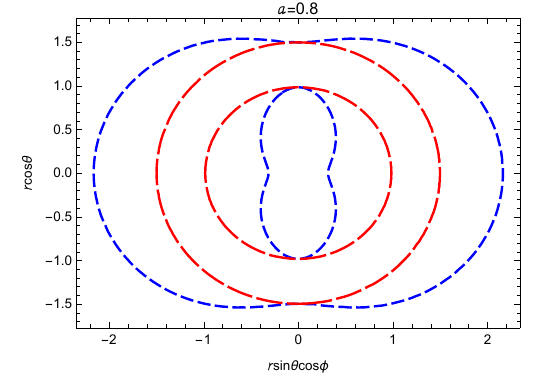}\hspace{-0.45cm}
			\includegraphics[scale=0.48]{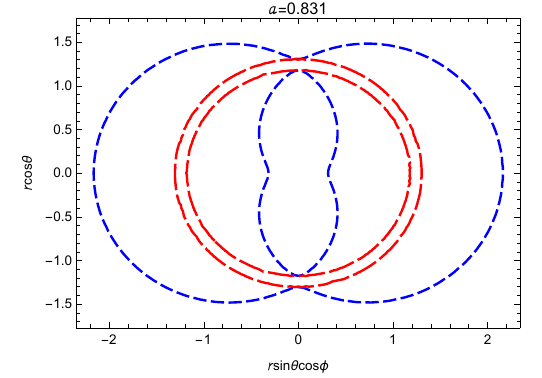}\hspace{-0.7cm}
			&\includegraphics[scale=0.48]{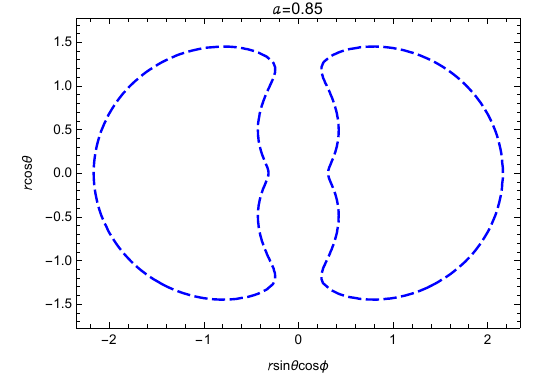}
		\end{tabular}
		\caption{The behavior of ergosphere in the xz-plane for different black hole parameters $q$, $b$, $\alpha$, $\omega _{q}$, $n$ and $a$.}
		\label{fig:ER}
	\end{center}
\end{figure*}

We know that the formula for redshift in a steady state spacetime is $\nu=\nu_{0} \sqrt{-g_{tt}}$, and we can find the infinite red-shift surface for $g_{tt}$=0. So for the infinite redshift surface of a rotating and twisting charged black holes with cloud of strings and quintessence, there is
\begin{equation}\label{eq:gtt}
	\begin{aligned}
		\begin{array}{c}
			\begin{aligned}
				g_{tt}=r^2-2 M r+a^2+q^2-n^2-\alpha r^{1-3 w}-b r^2-a^2\sin \theta=0.
			\end{aligned}
		\end{array}
	\end{aligned}
\end{equation}
From Eq.(\ref{eq:gtt}) it is possible to get two roots $r_{sls}^{+}$ and $r_{sls}^{-}$, where $r_{sls}^{+}$ is greater than $r_{sls}^{-}$, and $r_{sls}^{+}$ is the infinite red-shift surface. The region between the event horizon of the black hole $r_{H}^{+}$ and the outer infinite red-shift surface $r_{sls}^{+}$ is called the ergosphere, which corresponds to the outermost red curve to the blue curve in each graph in Fig.~\ref{fig:ER}.

We can obtain from Eqs.(\ref{eq:HO}) and (\ref{eq:gtt}) that the ergosphere of the black hole is directly affected by the parameters $q$, $b$, $\alpha$, $\omega _{q}$, $n$ and $a$. In Fig.~\ref{fig:ER}, we illustrate how the shape of the ergosphere is affected by black hole parameters.
It can be seen from Fig.~\ref{fig:ER} that the the ergosphere area increases with the increase of parameters $q$ and $a$, while decreases with the increase of parameters $b$, $\alpha$, $\omega _{q}$ and $n$.

\section{EQUATIONS OF MOTION AND THE CRITICAL ANGULAR MOMENTUM} \label{III}
To derive the CM energy for two particles, we first need to solve the equations of particle motion. In this section, we consider a time-like particle with a rest mass $m_0$ that falls from infinity on the equatorial plane ($\theta =\pi/2$, $u^\theta =0$) of the rotating and twisting charged black holes with cloud of strings and quintessence. The four-momentum of the particle is
\begin{equation}\label{eq:PT1}
	\begin{array}{c}
		\begin{aligned}
			P_{\mu} &=g_{\mu \nu} u^{\nu} ,
		\end{aligned}
	\end{array}
\end{equation}
where $u^{\nu}$ is the four-velocity of the particle.
From Eq.(\ref{eq:PT1}) and the energy and momentum conservation of the particle motion, we obtain
\begin{equation}\label{eq:PT}
	\begin{array}{c}
		\begin{aligned}
			P_{t} &=-E=g_{t t} u^{t}+g_{t \phi} u^{\phi  } ,
		\end{aligned}
	\end{array}
\end{equation}
\begin{equation}\label{eq:PF}
	\begin{array}{c}
		\begin{aligned}
			P_{\phi} &=L=g_{\phi \phi} u^{\phi  }+g_{t \phi} u^{t},
		\end{aligned}
	\end{array}
\end{equation}
where $E$ and $L$ are the energy of the particle per unit mass and the angular momentum of the particle per unit mass parallel to the rotation axis of the black hole, respectively. The four-velocity of the particle is simplified from Eqs.(\ref{eq:PT}), (\ref{eq:PF}) and $P_{\nu} P^{\nu}=-m_{0}^{2}$ as
\begin{figure*}
	\begin{tabular}{c c}
		\includegraphics[scale=0.5]{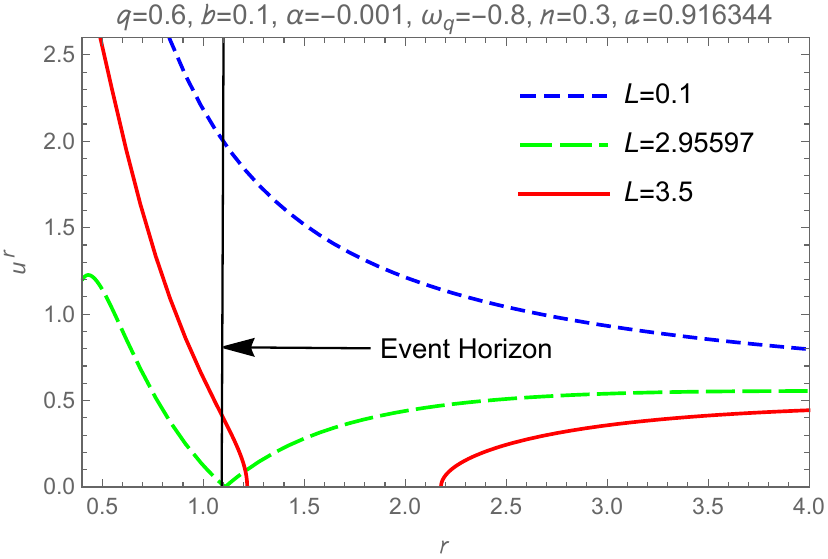}\hspace{-0.2cm}
		&\includegraphics[scale=0.5]{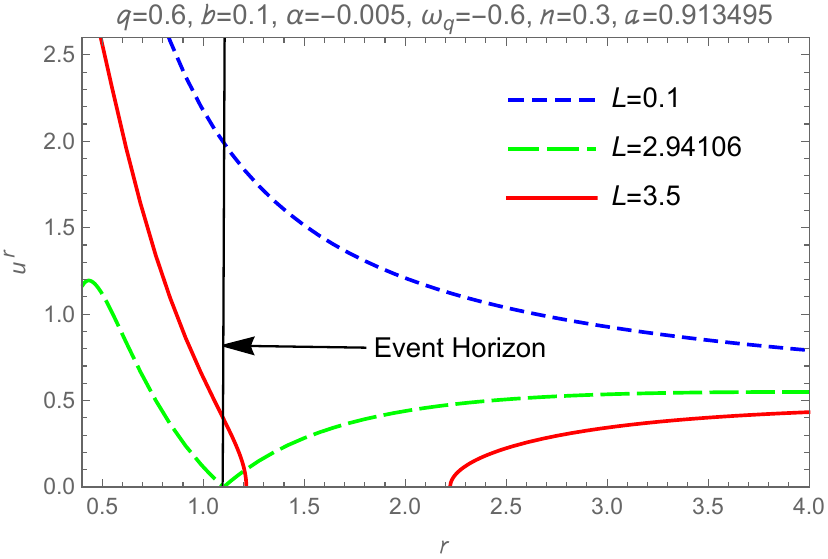}\\
		\includegraphics[scale=0.5]{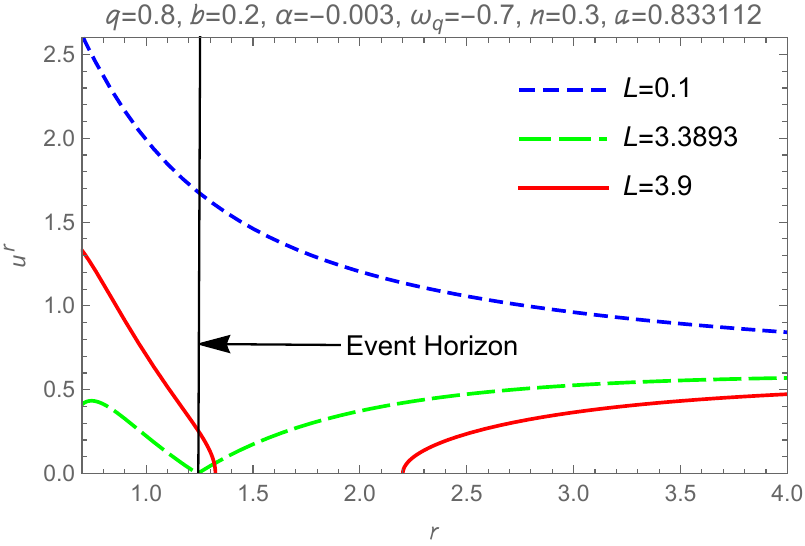}\hspace{-0.2cm}
		&\includegraphics[scale=0.5]{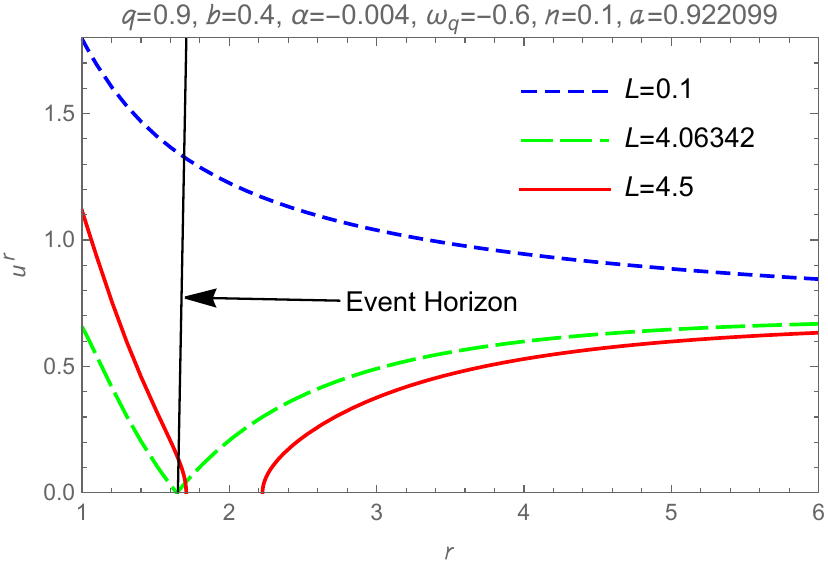}
	\end{tabular}
	\caption{The behavior of $u^r$  vs $r$ for extremal black hole.}\label{fig:dotr}
\end{figure*}
\begin{equation}\label{eq:T}
	\begin{array}{c}
		\begin{aligned}
			u^{t} &=\frac{Z^2-\Delta Y^2}{\Delta T}E+\frac{\Delta Y-aZ}{\Delta T}L,
		\end{aligned}
	\end{array}
\end{equation}
\begin{equation}\label{eq:F}
	\begin{array}{c}
		\begin{aligned}
			u^{\phi  } &=\frac{aZ-\Delta Y}{\Delta T}E+\frac{\Delta-a^2}{\Delta T}L,
		\end{aligned}
	\end{array}
\end{equation}
\begin{equation}\label{eq:R}
	\begin{array}{c}
		\begin{aligned}
			u^{r} &=\pm\frac{\sqrt{(Z^2-\Delta Y^2)E^2
					-(\Delta-a^2)L^2+2(\Delta Y-aZ)EL- m_0^2 \Delta T } }{T} ,
		\end{aligned}
	\end{array}
\end{equation}
here $Z=r^2+(a+n)^2$, $Y=a+2n$ and $T=r^2+n^2$. The plus and minus signs in Eq.(\ref{eq:R}) correspond to the outgoing and incoming geodesics, respectively. We know that for a particle with a certain angular momentum, it must satisfy a certain range of angular momentum values to be able to fall into a black hole. If not, the particle cannot fall into a black hole. In order to obtain the angular momentum range of the particle entering the black hole, we consider the relationship between the radial velocity of the particle and the effective potential as follows:
\begin{equation}\label{eq:V1}
	\begin{array}{c}
		\begin{aligned}
			\frac{1}{2} (u^{r})^{2}+V_{{eff }}=0.
		\end{aligned}
	\end{array}
\end{equation}
Substitute Eq.(\ref{eq:R}) into Eq.(\ref{eq:V1}) to get the effective potential as
\begin{equation}\label{eq:V2}
	\begin{array}{c}
		\begin{aligned}
			V_{eff}=-\frac{(Z^2-\Delta Y^2)E^2
				-(\Delta-a^2)L^2+2(\Delta Y-aZ)EL-\Delta T }{2T^2}.
		\end{aligned}
	\end{array}
\end{equation}

\begin{figure*}
	\begin{tabular}{c c}
		\includegraphics[scale=0.45]{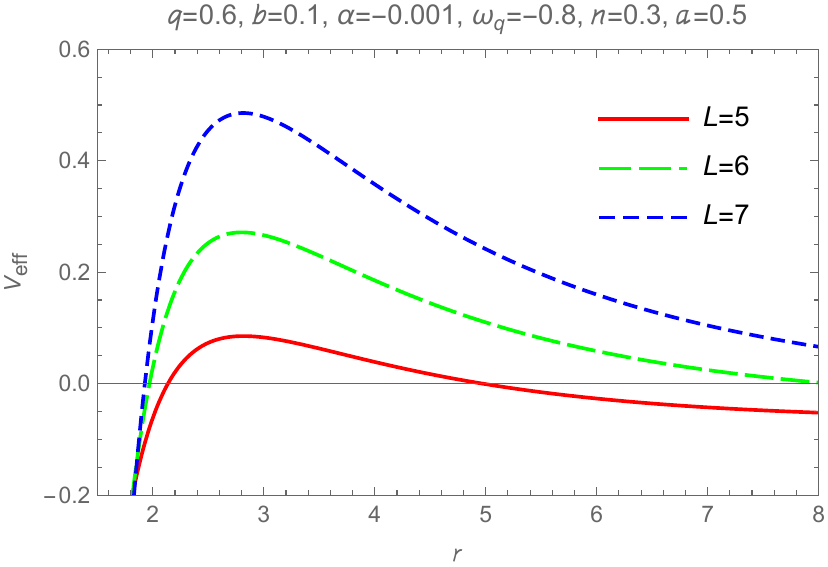}\hspace{-0.2cm}
		&\includegraphics[scale=0.45]{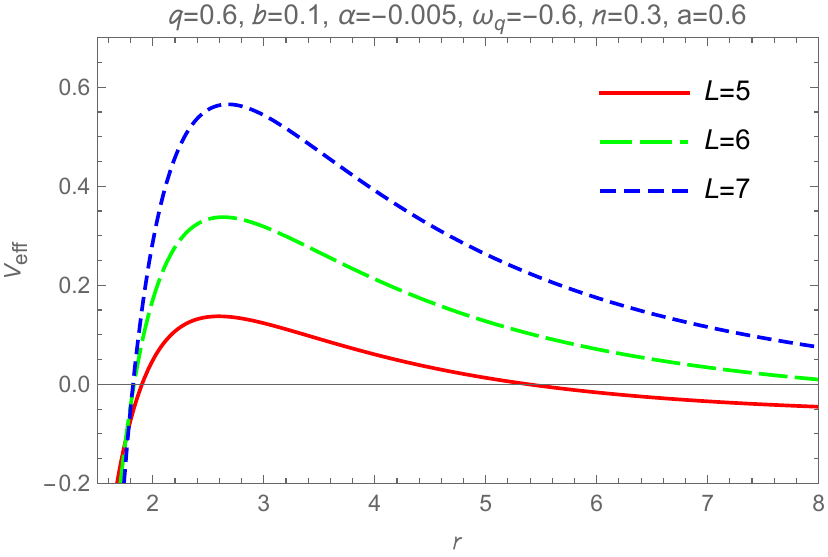}\\
		\includegraphics[scale=0.45]{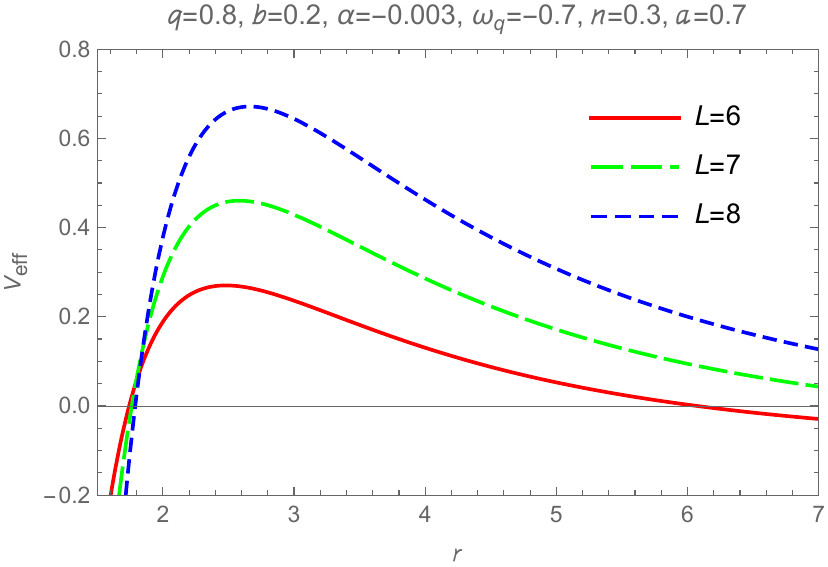}\hspace{-0.2cm}
		&\includegraphics[scale=0.45]{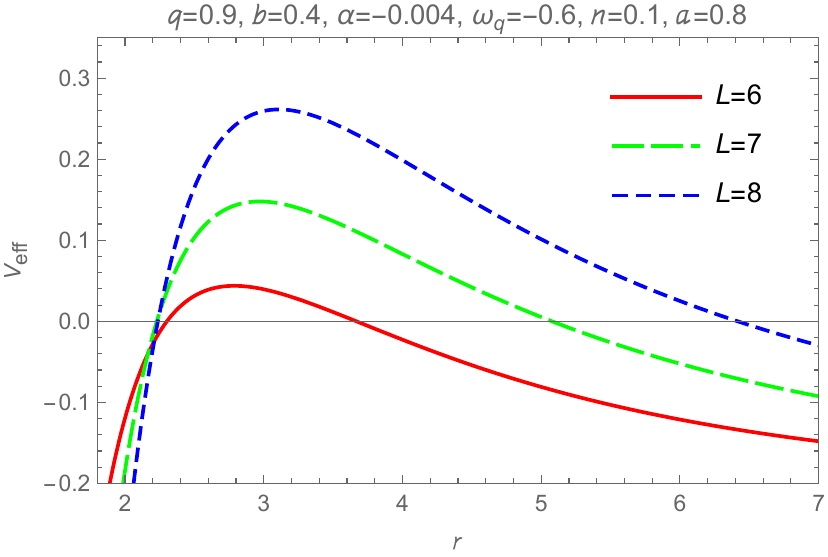}\\
	\end{tabular}
	\caption{The behavior of $V_{eff}$ vs $r$ for non-extreme black hole.}\label{fig:Veff}
\end{figure*}
The condition that the motion trajectory of a particle is a circular orbit is as follows:
\begin{equation}\label{eq:L}
	\begin{array}{c}
		\begin{aligned}
			V_{e f f}=0 \quad \text { and } \quad \frac{d V_{e f f}}{d r}=0.
		\end{aligned}
	\end{array}
\end{equation}
From the above condition, we obtained Tables \ref{table1} and \ref{table2} by numerical calculation method. They correspond to the range of values of the angular momentum of the particle entering the black hole under different parameters for extreme and non-extreme black holes, respectively. For a time like particle, the geodesic satisfies
\begin{equation}\label{eq:T1}
	\begin{array}{c}
		\begin{aligned}
			(Z^2-\Delta Y^2)E+(\Delta Y-aZ)L>0.
		\end{aligned}
	\end{array}
\end{equation}
When $r$ approaches $r^+_H$, the above equation reduces to
\begin{equation}\label{eq:L1}
	\begin{array}{c}
		\begin{aligned}
			E-\Omega_{H} L \geq 0, 	   \Omega_{H}=\frac{a}{r_{H}^{+}+(a+n)^{2}},
		\end{aligned}
	\end{array}
\end{equation}
where $\Omega_{H}$ is the general expression of the horizon angular velocity of a black hole with rotation parameter and twisting parameter. The critical angular momentum $L_c =E/\Omega_{H}$ is obtained when we take the equal sign of Eq.(\ref{eq:L1}).

\begin{table}
	\begin{center}
		\caption{Limit values of angular momentum for different extreme values of rotating and twisting charged black holes with cloud of strings and quintessence.}\label{table1}
		\begin{tabular}{| c c | c | c | c | c | c |c | c | c |}
			\hline
			& $q$	&$b$     &$\alpha$    &$\omega _{q}$    &$n$      & $a_{E}$      & $r^{E}_{H}$     & $L_{1}$     & $L_{2}$  \\
			\hline
			& 0 	& 0         & 0          & 0     	      &0       & 1             & 1.0           & 2          &-4.82843  \\
			& 0.6 	& 0.1    & -0.001      & -0.8  	        &0.4       & 0.953774      & 1.10869       & 3.21030      &-5.08470  \\
			& 0.6 	& 0.1    & -0.005      & -0.6 	        &0.3        & 0.913495      & 1.10185       & 2.94106    &-5.07633  \\
			& 0.7  	& 0.2     & -0.002     & -0.7		    &0.4       & 0.957098      & 1.24387       & 3.54084    &-6.51537  \\
			& 0.7 	& 0.2    & -0.003      &-0.8 	       	&0.3        & 0.918524      & 1.23933       & 3.28869    &-6.25157  \\
			& 0.8 	& 0.1    & -0.004      & -0.8     	    &0.4        &0.790868      & 1.10158        & 3.32754    &-4.37907  \\
			& 0.8 	& 0.2     & -0.003     & -0.7           &0.3        & 0.833112      & 1.24086       & 3.38930    &-6.28568  \\
			& 0.9 	& 0.3    & -0.002      & -0.5       	&0.2        & 0.808529      & 1.42251       & 3.76075    &-8.45353  \\
			& 0.9 	& 0.4    &-0.004       & -0.6 		    &0.1        & 0.922099      & 1.64383       & 4.06342    &-11.2416  \\
			& 0.9  	& 0.3     & -0.006     & -2/3		    &0.5        & 0.922791      & 1.40325       & 4.32758    &-7.63874  \\
			
			\hline
		\end{tabular}
	\end{center}
\end{table}
 In Figs.~\ref{fig:dotr} and \ref{fig:Veff}, we show the behavior of $u^{r}$ with radius $r$ for the extreme black hole and the behavior of $V_{eff}$ with radius $r$ for the non-extreme black hole, respectively. From the above calculation, Figs.~\ref{fig:dotr},~\ref{fig:Veff} and Tables \ref{table1}, \ref{table2}, we get the following conclusions:
\begin{table}
	\begin{center}
		\caption{The limiting values of angular momentum for different non-extremal cases of rotating and twisting charged black holes with cloud of strings and quintessence.}\label{table2}
		\begin{tabular}{| c c | c | c | c | c | c |c | c | c | c |}
			\hline
			& $q$	&$b$      &$\alpha$    &$\omega _{q}$  &$n$       & $a$     & $r^{-}_{H}$    & $r^{+}_{H}$  & $L_{1}$     & $L_{2}$  \\
			\hline
			& 0 	& 0       & 0          & 0     	      &0          & 0.8        & 0.4           & 1.6         & 2.89443      &-4.68328  \\
			& 0.6 	& 0.1    & -0.001      & -0.8  	      &0.4        & 0.7      & 0.427123       & 1.78925       &4.35283      &-4.81752  \\
			& 0.6 	& 0.1    & -0.005      & -0.6 	      &0.3         & 0.8      & 0.640087       & 1.56284      & 3.68998      &-4.96037  \\
			& 0.7  	& 0.2     & -0.002     & -0.7	      &0.4         & 0.7      & 0.517077       & 1.96908      & 4.96519     &-6.13192  \\
			& 0.7 	& 0.2    & -0.003      &-0.8 	    	&0.4        & 0.8      & 0.659161       & 1.81695      & 4.58495     &-5.95648  \\
			& 0.8 	& 0.1    & -0.004      & -0.8     	    &0.4        &0.7      & 0.716958        & 1.48493     & 3.90718      &-4.31098  \\
			& 0.8 	& 0.2     & -0.003     & -0.7           &0.3        & 0.8      & 0.982703       & 1.49871      & 3.84260     &-6.23697  \\
			& 0.9 	& 0.3    & -0.002      & -0.5       	&0.2        & 0.7      & 0.940344       & 1.90451      & 4.84210    &-8.21122  \\
			& 0.9 	& 0.4    &-0.004       & -0.6 		    &0.1        & 0.8      & 1.05842       & 2.22789      & 5.54924      &-10.9108  \\
			& 0.9  	& 0.3     & -0.006     & -2/3		    &0.5        & 0.7      & 0.695129       & 2.10725     & 5.83798      &-7.23615  \\
			\hline
		\end{tabular}
	\end{center}
\end{table}

\begin{itemize} 	
	\item{Fig.~\ref{fig:dotr} shows that for the extreme black hole, when the particle reaches the event horizon with angular momentum $L=L_{C}$, its radial velocity becomes zero. The particle will fall into the black hole if its angular momentum is $L<L_{C}$. The particle will not fall into the black hole if its angular momentum is $L>L_{C}$.}
	
	\item{As shown in Table \ref{table1}, the maximum value of the angular momentum range of particles incident into the extreme black hole is exactly the value of critical angular momentum and the radius of orbit is the radius of the event horizon. As shown in Table \ref{table2}, the critical angular momentum for non-extreme black holes does not meet the range of angular momentum that particles can fall into the black hole.}
	\item {There are three conclusions for the critical angular momentum when determining the parameters of extreme black holes.}
	\begin{enumerate}
		\item[1)] The critical angular momentum can be obtained from Eq.(\ref{eq:R}) by setting $r_{H}^{+}$ and $u^r=0$.
		\item[2)] The critical angular momentum can be obtained from the maximum value($L_{1}$ in Table \ref{table1}) of the particle incident angular momentum range.
		\item[3)] The critical angular momentum can be obtained by substituting the radius of the event horizon into Eq.(\ref{eq:L1}). This method can be directly applied to any black hole with parameters $a$ and $n$, and the critical angular momentum can be obtained immediately.
	\end{enumerate}
\end{itemize}

\section{NEAR HORIZON COLLISION IN ROTATING AND TWISTING CHARGED BLACK HOLES WITH CLOUD OF STRINGS AND QUINTESSENCE}\label{IV}
In this section, we will use the four-velocities results of the previous section to calculate the CM energy expression of two particles with different masses outside the black hole, and discuss whether the CM energy can diverge when two particles collide at the event horizon under extreme and non-extreme black holes, and if so, what are the conditions. We examine the case of two particles with rest masses $m_1$ and $m_2$, angular momenta $L_1$ and $L_2$, and energies $E_1=E_2=1$, falling from infinity on the equatorial plane of the rotating and twisting charged black holes with cloud of strings and quintessence. The four-momentum expression for a particle is given by
\begin{equation}\label{eq:P}
	\begin{array}{c}
		\begin{aligned}
			p_{i}^{\mu}=m_{i} u_{i}^{\mu},
		\end{aligned}
	\end{array}
\end{equation}
where $i=1, 2$, $m_i$ and $u_{i}^{\mu}$ correspond to the rest mass and four-velocities of the $i$ particle. The total four-momentum of two particles is
\begin{equation}\label{eq:P1}
	\begin{array}{c}
		\begin{aligned}
			P_{t}^{\mu}=P_{(1)}^{\mu}+P_{(2)}^{\mu}.
		\end{aligned}
	\end{array}
\end{equation}
The CM energy of the two particles can be expressed as
\begin{equation}\label{eq:E1}
	\begin{array}{c}
		\begin{aligned}
			E_{C M}^{2}&=-P_{t}^{\mu} P_{t\mu}=-\left(P_{(1)}^{\mu}+P_{(2)}^{\mu}\right)\left(P_{(1) \mu}+P_{(2) \mu}\right) \\
			&=-\left(m_{1} u_{(1)}^{\mu}+m_{2} u_{(2)}^{\mu}\right)\left(m_{1} u_{(1) \mu}+m_{2} u_{(2) \mu}\right) .
		\end{aligned}
	\end{array}
\end{equation}
By substituting the normalization condition for the four-velocity $u^{\nu}u_{\nu}=-1$ into the above equation, it can be simplified as \cite{Harada:2010yv}
\begin{equation}\label{eq:E2}
	\begin{array}{c}
		\begin{aligned}
			\frac{E_{C M}^{2}}{2 m_{1} m_{2}}=\frac{\left(m_{1}-m_{2}\right)^{2}}{2 m_{1} m_{2}}+1-g_{\mu \nu} u_{(1)}^{\mu} u_{(2)}^{\nu}.
			
		\end{aligned}
	\end{array}
\end{equation}
By substituting Eqs.(\ref{eq:T}), (\ref{eq:F}) and (\ref{eq:R}) into the Eq.(\ref{eq:E2}), we obtain
\begin{equation}\label{eq:E3}
	\begin{array}{c}
		\begin{aligned}
			\frac{E_{C M}^{2}}{2 m_{1} m_{2}}&=\frac{\left(m_{1}-m_{2}\right)^{2}}{2 m_{1} m_{2}}+\frac{1 }{\Delta T}(\Delta T+(Z^2-\Delta Y^2)\\
			&-(\Delta-a^2)L_1L_2+(\Delta Y-aZ)(L_1+L_2)\\
			&-\sqrt{-\Delta T m_1^2 +(Z^2-\Delta Y^2)-(\Delta-a^2)L_1^2+2(\Delta Y-aZ)L_1} \\
			&\quad \sqrt{-\Delta T m_2^2 +(Z^2-\Delta Y^2)-(\Delta-a^2)L_2^2+2(\Delta Y-aZ)L_2}) .
		\end{aligned}
	\end{array}
\end{equation}
From this formula, it can be seen that the CM energy of the two particles outside the black hole is not only related to the mass and angular momentum of the two particles, but also affected by the black hole parameters $q$, $b$, $\alpha$, $\omega _{q}$, $n$ and $a$. For a real black hole, its parameters are all determined. To use it as a particle accelerator, we can only adjust the energy level of the black hole as a particle accelerator by changing the mass and angular momentum of the particles, and the angular momentum of the particles has a more significant impact on the adjustment of the CM energy. Next, we will discuss the case of extreme and non extreme black holes as particle accelerator by adjusting the angular momentum of particles.
\begin{figure*}
	\begin{center}
		\begin{tabular}{c c c c}
			\includegraphics[scale=0.6]{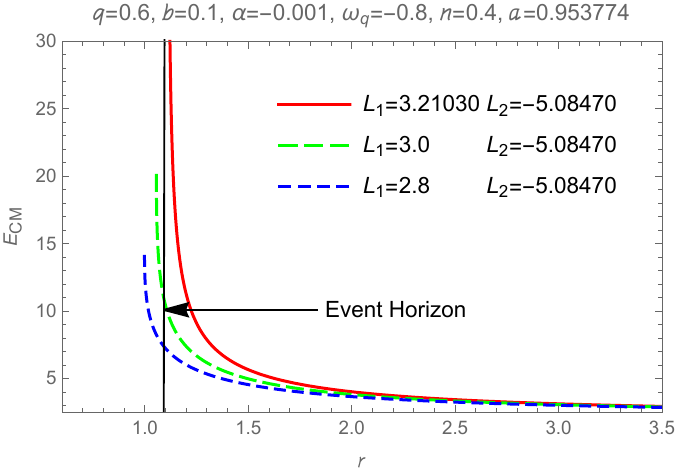}\hspace{-0.2cm}
			&\includegraphics[scale=0.6]{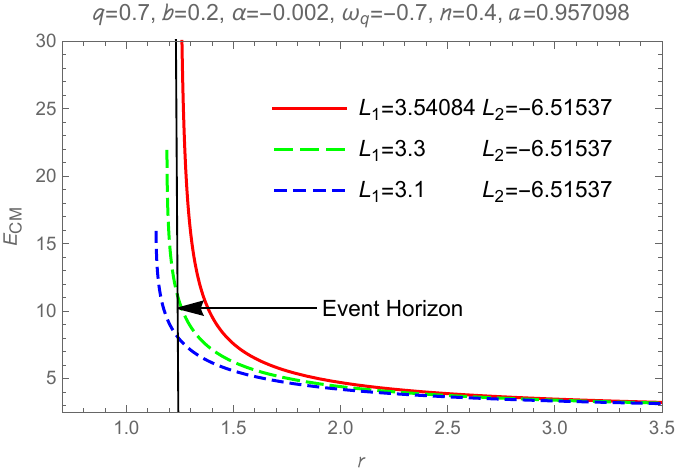}\\
			\includegraphics[scale=0.6]{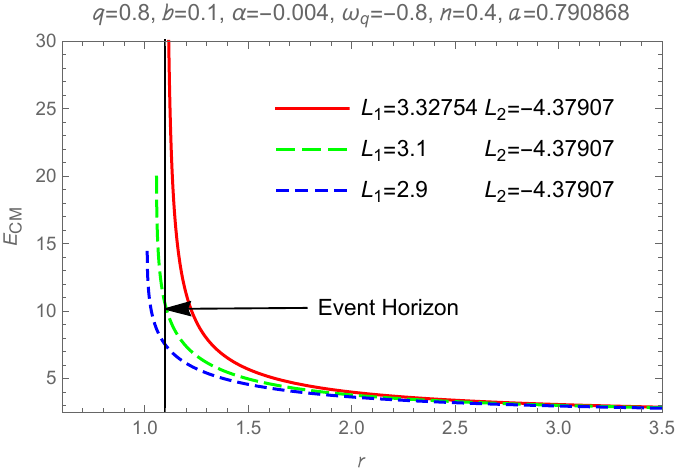}\hspace{-0.2cm}
			&\includegraphics[scale=0.6]{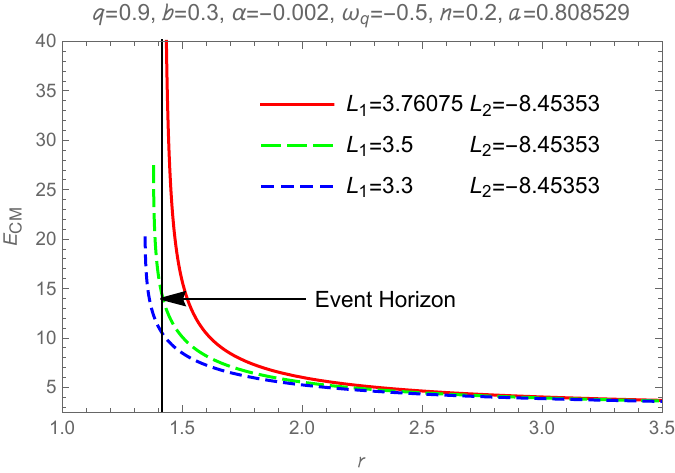}
		\end{tabular}
		\caption{The behavior of $E_{CM}$ vs $r$ for extremal black hole.}\label{fig:Ecm1}
	\end{center}
\end{figure*}

For extreme rotating and twisting charged black holes with cloud of strings and quintessence, when the black hole parameters are determined (fixed $m_1=m_2=1$, $q=0.8$, $b=0.1$, $\alpha=-0.004$, $\omega _{q}=-0.8$, $n=0.4$ and $a_E=0.790868$). We can get the following expression by taking the limit as $r$ approaches $r^E_H$ of Eq.(\ref{eq:E3}):
\begin{equation}\label{eq:E4}
	\begin{array}{c}
		\begin{aligned}
			\lim_{r \to r^E_H}\frac{E_{C M}^{2}}{2} &=7.17544 + 0.531499 (L_1+L_2)-0.72808 L_1L_2\\&-\frac{B}{(L_1-L_C)(L_2-L_C)}
			+\frac{A_1 (L_2-L_C)}{(L_1-L_C)} +\frac{A_2 (L_1-L_C)}{(L_2-L_C)},
		\end{aligned}
	\end{array}
\end{equation}
where
\begin{equation}\label{eq:E5}
	\begin{array}{c}
		\begin{aligned}
			A_i&=-11.7599-2.34076 L_i+11.0376 L_i^2-4.72318 L_i^3+0.582025 L_i^4 ,(i=1,2) \\
			B&=-26.6415+8.00637 (L_1+L_2)-2.40609 L_2 L_1 , r^E_H=1.10158.
		\end{aligned}
	\end{array}
\end{equation}
The above formula shows that if two particles approach the event horizon and one of them has critical angular momentum $L_{C} =\frac{E(r_{H}^{E}+(a+n)^{2})}{a}=3.32754$, then the CM energy diverges. The critical angular momentum $L_{C}$ of the particle satisfies Table \ref{table1}, which means that the particle can reach the black hole horizon.
Fig.~\ref{fig:Ecm1} shows the behavior of $E_{CM}$ vs $r$ for extreme black holes with different parameters $q$, $b$, $\alpha$, $\omega _{q}$, $n$ and $a$. For an extremal black hole as a particle accelerator, Eq.(\ref{eq:E4}) and Fig.~\ref{fig:Ecm1} show that the CM energy of any energy level can be obtained at the event horizon by adjusting the angular momentum of particles.
\begin{figure*}
	\begin{center}
		\begin{tabular}{c c c c}
			\includegraphics[scale=0.55]{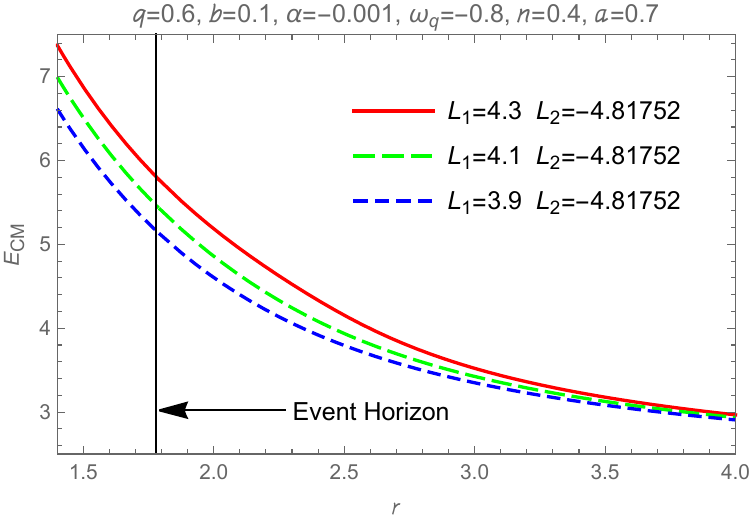}\hspace{-0.2cm}
			\includegraphics[scale=0.55]{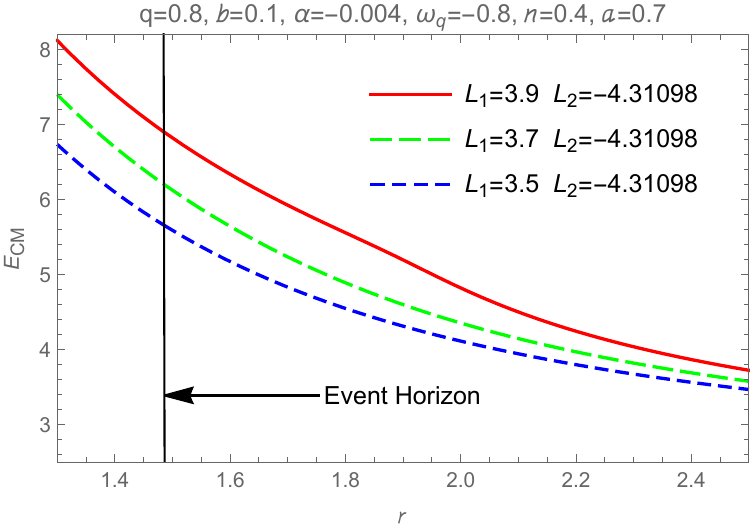}\hspace{-0.2cm}
		\end{tabular}
		\caption{The behavior of $E_{CM}$ vs $r$ for non-extremal black hole.}\label{fig:Ecm2}
	\end{center}
\end{figure*}

For non-extreme rotating and twisting charged black holes with cloud of strings and quintessence, when the black hole parameters are determined (fixed $m_1=m_2=1$, $q=0.8$, $b=0.1$, $\alpha=-0.004$, $\omega _{q}=-0.8$, $n=0.4$ and $a=0.7$), we can get the following expression by taking the limit as $r$ approaches $r^+_H$ of Eq.(\ref{eq:E3}):
\begin{equation}\label{eq:E6}
	\begin{array}{c}
		\begin{aligned}
			\lim_{r \to r^+_H}\frac{E_{C M}^{2}}{2} &=12.1628-0.607315 (L_1+L_2)
			-0.422832 L_2 L_1+\frac{C}{(L_1-L_C)(L_2-L_C)},
		\end{aligned}
	\end{array}
\end{equation}
where
\begin{equation}\label{eq:E7}
	\begin{array}{c}
		\begin{aligned}
			C&=-118.522+31.3771 (L_1+L_2)-5.80718 L_2 L_1-0.713202 (L_1 L_2^2+L_2 L_1^2)\\&-0.0242909 (L_2^2+L_1^2)+0.207189 L_1^2 L_2^2, r^+_H=1.48493 .
		\end{aligned}
	\end{array}
\end{equation}
For non-extreme black holes, the above formula suggests that the CM energy diverges like in extreme black holes if two particles approach the event horizon with one having a critical angular momentum $L_{C} =\frac{E(r_{H}^{+}+(a+n)^{2})}{a}=4.87896$, but this is impossible. Because the critical angular momentum in a non-extreme black hole does not satisfy the angular momentum range ($-4.31098\sim 3.90718$) of the particle incident into the black hole, the particle cannot reach the event horizon. Fig.~\ref{fig:Ecm2} shows the behavior of $E_{CM}$ vs $r$ for non-extreme black holes with different parameters $q$, $b$, $\alpha$, $\omega _{q}$, $n$ and $a$.
From Eq.(\ref{eq:E6}) and Fig.\ref{fig:Ecm2}, it is known that it is impossible to make the CM energy of two particles diverge at the event horizon of the black hole by adjusting the particle angular momentum, and the obtained CM energy  should have a maximum value.

\section{CONCLUSION} \label{V}
In this paper, we have investigated the event horizon and ergosphere of rotating and twisting charged black holes with cloud of strings and quintessence. We have also examined the CM energy of particles for extreme and non-extreme black holes as particle accelerator, respectively. The event horizon of the black hole is affected by the parameters $q$, $b$, $\alpha$, $\omega _{q}$, $n$ and $a$. When the other parameters are fixed, for the parameters $q$ and $a$, the distance between the inner and outer horizons of the black hole decreases with the increase of the parameter value; when $q$ or $a$ is equal to critical value, the inner and outer horizons coincide; when $q$ or $a$ is greater than the critical value, there is no black hole. For the parameter $n$, the distance between the inner and outer horizons of the black hole decreases with the decrease of the parameter value; when $n$ is equal to critical value, the inner and outer horizons coincide; when $n$ is less than the critical value, there is no black hole. Both the minimum of the $\Delta$ function and the radial coordinate $r_{min}$ corresponding to the minimum of the $\Delta$ function decrease as the values of parameters $\alpha$, $\omega _{q}$ and $b$ increase. Parameter $\alpha$ when $\omega _{q}=-1/3$ has a similar effect on the $\Delta$ function as parameter $b$ when $\omega _{q}$ is equal to other values. The ergosphere area is affected by the black hole parameters, where the parameters $q$ and $a$ enlarge the ergosphere region, while the parameters $b$, $\alpha$, $\omega _{q}$ and $n$ shrink it.

We have calculated the equations of particle motion on the black hole equatorial plane and obtained the range of angular momentum (Table \ref{table1} and \ref{table2}) of particles falling into the black hole. We have introduced the critical angular momentum $L_c$, which is also the maximum value $L_1$ of the angular momentum range (Table \ref{table1}). We have found that a particle with angular momentum $L=L_C$ falling from infinity can just reach the event horizon of the black hole and satisfy the circular orbit condition Eq.(\ref{eq:R}) in the extreme black hole background. For non-extreme black holes, the critical angular momentum does not meet Table \ref{table2}. Therefore, the particle cannot reach the event horizon of the black hole and there is no result similar to the motion of particle of extreme black holes.

We have obtained the CM energy expression of two particles with different masses from the equations of particle motion, and found that the CM energy of the particle is related to the parameters of the black hole and the parameters of the particle itself. We have got the CM energy expression of two particles at the event horizon of an extreme and a non-extreme black hole by Eq.(\ref{eq:E3}). It has been found that for both extreme and non-extreme black holes, when one of the two particles has angular momentum $L=L_C$, the CM energy diverges. But from the above discussion on the angular momentum range of the particle falling into the black hole, we know that $L_C$ does not meet Table \ref{table2} for non-extreme black holes, that is, we cannot obtain the above results for non-extreme black holes. Thus, we have proved that for extreme black holes, we can adjust the angular momentum of the incident particles to make the particle CM energy has any magnitude when the particles reach the event horizon of the black hole. For non-extreme black holes, adjusting the angular momentum of the incident particles makes the particle CM energy has a maximum value but not diverges when the particles reach the event horizon of the black hole.

\end{document}